# Multiple magnetic transitions and magnetocaloric effect in $Gd_{1-x}Sm_xMn_2Ge_2$ compounds

**Pramod Kumar[1], Niraj K. Singh[1], K.G. Suresh[1] (\*), A.K. Nigam[2] and S.K. Malik[2]**
[1]Department of Physics, I. I. T .Bombay, Mumbai 400076, India
[2]Tata Institute of Fundamental Research, Homi Bhabha Road, Mumbai 400005, India

**Abstract**.-Magnetic and magnetocaloric properties of polycrystalline samples of $Gd_{1-x}Sm_xMn_2Ge_2$ have been studied. All the compounds except $GdMn_2Ge_2$ show re-entrant ferromagnetic behavior. Multiple magnetic transitions observed in these compounds are explained on the basis of the temperature dependences of the exchange strengths of the rare earth and Mn sublattices. Magnetocaloric effect is found to be positive at the re-entrant ferromagnetic transition, whereas it is negative at the antiferro-ferromagnetic transition. In $SmMn_2Ge_2$, the magnetic entropy change associated with the re-entrant transition is found to decrease with field, which is attributed to the admixture effect of the crystal field levels. The isothermal magnetic entropy change is found to decrease with increase in Sm concentration.

………………………………………………………………………………………………..

(\*) E-mail:suresh@phy.iitb.ac.in

The magnetocaloric effect (MCE) was discovered in 1881 by Warburg [1]. It is the response of a magnetic solid to a changing magnetic field, which manifests as a change in its temperature. For a simple ferromagnet near its Curie temperature ($T_c$), when a magnetic field is applied the moments get aligned parallel to the magnetic field, which lowers the magnetic entropy and causes the sample to heat up. When the magnetic field is turned off, the moments tend to randomize, thereby increasing the entropy and the material cools off. This process has been used for achieving sub-Kelvin temperatures and this technique is known as adiabatic demagnetization. Recently, this technique is being considered for cryogenic applications in sub-room temperature and near room temperature regimes [2-5].

The search for potential working substances suitable for cooling to the sub-room temperature and near room temperature regime is essential for further improvements in the field of magnetic refrigeration. Materials with first order magnetic transition are of particular interest, since they exhibit significant changes in the isothermal magnetic entropy ($\Delta S_M$) and the adiabatic temperature change ($\Delta T_{ad}$) at the magnetic transition temperature. Compounds showing field-induced magnetic transitions and/or structural transitions have been found to exhibit very large values of MCE, as in the case of $Gd_5Si_2Ge_2$ [6]. Giant magnetocaloric effect was also found in systems such as MnAs, $LaFe_{11.4}Si_{1.6}$ [7-10], by virtue of a first-order ferromagnetic to paramagnetic transition.

Among the rare earth (R) – transition metal intermetallics, $RMn_2Ge_2$ have attracted a lot of attention due to their interesting magnetic properties [11-14]. These compounds, in general, crystallize in the tetragonal $ThCr_2Si_2$-type structure. In this series of compounds, both R and Mn atoms possess magnetic moments and form a layered structure of Mn-Ge-R-Ge-Mn type, with the layers perpendicular to the c-axis. The inter-layer and intra-layer Mn-Mn exchange interactions are very sensitive to the intra-layer Mn distance, leading to ferromagnetic or antiferromagnetic ordering of the Mn sublattice. In most of the $RMn_2Ge_2$ compounds, the intra-layer Mn sublattice, in general, shows the antiferromagnetic ordering above the room temperature while R sublattice remains disordered. As temperature is reduced, the R moments also order magnetically and this causes a change in the magnetic ordering of the Mn sublattice. By virtue of the different temperature dependences of R and Mn sublattice magnetizations, these compounds, in general, shows multiple magnetic transitions, re-entrant ferromagnetism etc. Furthermore, the compounds with rare earths such as Sm, Gd, Dy and Tb show first order magnetic transitions (FOT) [15]. The occurrence of FOT is expected to result in considerable MCE.

Recently, as part of investigations on the magnetic and magnetocaloric properties, we have reported our results on $Gd_{1-x}Sm_xMn_2Si_2$ [16] and $GdMn_2Si_{2-x}Ge_x$ [17] compounds. It is known that the magnetic nature of $SmMn_2Ge_2$ and $GdMn_2Ge_2$ shows some difference on account of their lattice parameter variations [18]. Therefore, a solid solution of these two compounds, i.e. $Gd_{1-x}Sm_xMn_2Ge_2$ would be interesting from the point of magnetic properties as well as other related properties like MCE. With this aim, we have studied the effect of Sm substitution for Gd on the magnetic and magnetocaloric properties of $Gd_{1-x}Sm_xMn_2Ge_2$ with x=0, 0.4, 0.6 and 1 and the results are presented in this paper.

Polycrystalline samples of $Gd_{1-x}Sm_xMn_2Ge_2$ with x=0, 0.4, 0.6 and 1 were synthesized by arc melting the constituent elements in stochiometric proportions in a water-cooled copper hearth under high pure argon atmosphere. The purity of the starting elements was 99.9% for the rare earths and 99.99 % for Mn and Ge. The ingots were melted several times to ensure homogeneity. The as-cast samples were characterized by room temperature powder x-ray diffractograms collected using Cu-$K_\alpha$ radiation. The magnetization (M) measurements, both under 'zero-field-cooled' and 'field-cooled' conditions, in the temperature (T) range of 5-300 K and up to a maximum field (H) of 80 kOe were performed using a vibrating sample magnetometer (VSM, OXFORD instruments). The temperature variation of magnetization was measured in a field of 200 Oe, in the warming cycle.

The Rietveld refinement of the x-ray diffractograms collected at room temperature confirms that all the compounds have formed in single phase with the $ThCr_2Si_2$ structure (Space group=I4/mmm). The lattice parameters and the Mn-Mn bond lengths along the a-axis and the c-axis (i.e. $d^a_{Mn-Mn}$ and $d^c_{Mn-Mn}$), as obtained from the refinement, are given in Table I. It may be noticed from the table that $d^a_{Mn-Mn}$ increases with increase in the Sm concentration, whereas $d^c_{Mn-Mn}$ remains almost a constant. This increase in $d^a_{Mn-Mn}$ is attributed to the larger ionic radius of Sm, compared to that of Gd. It can also be noticed that in all the compounds, the intra-layer Mn-Mn distance ($d^a_{Mn-Mn}$) is greater than the critical distance of 2.85 Å and therefore, the inter-layer magnetic coupling is ferromagntic.

Table 1 Lattice parameters, bond lengths and the magnetic transition temperatures in $Gd_{1-x}Sm_xMn_2Ge_2$ compounds.

| x | a (Å) | c (Å) | $d^a_{Mn-Mn}$ (Å) | $d^c_{Mn-Mn}$ (Å) | $T_1$ (K) | $T_2$ (K) |
|---|---|---|---|---|---|---|
| 0 | 4.026 | 10.881 | 2.847 | 5.441 | 96 | --- |

| 0.4 | 4.043 | 10.884 | 2.859 | 5.442 | 85 | 296 |
| 0.6 | 4.051 | 10.888 | 2.864 | 5.444 | 84 | 210 |
| 1 | 4.062 | 10.889 | 2.872 | 5.445 | 110 | 146 |

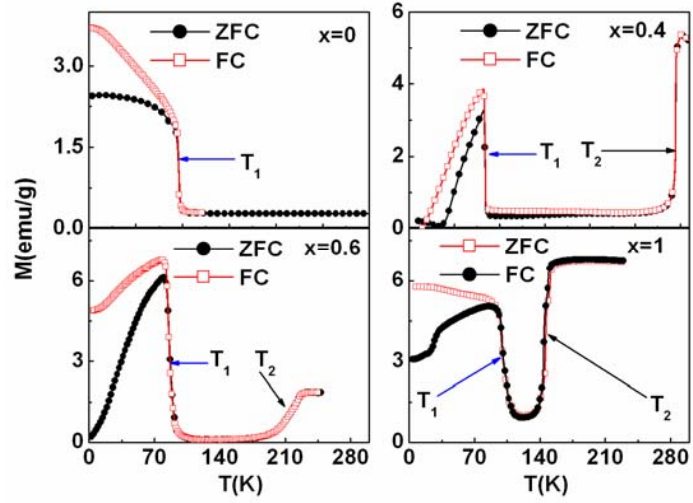

Fig.1 Temperature dependence of magnetization in $Gd_{1-x}Sm_xMn_2Ge_2$ compounds in a field of 200 Oe. The filled circles show the ZFC data and the open circles show the FC data.

Fig.1 shows the temperature variation of magnetization of $Gd_{1-x}Sm_xMn_2Ge_2$ compounds obtained in a field of 200 Oe. As can be seen, there are two transitions, one at $T_1$ and the other at $T_2$ in all the compounds, except $GdMn_2Ge_2$. In the case of $GdMn_2Ge_2$, there is only one transition at $T_1=96$ K. In this case, as is evident from Table 1, $d^a_{Mn-Mn}$ is slightly less than the critical distance needed for the ferromagnetic coupling of the inter-layer Mn-Mn moments. On the other hand, the $d^a_{Mn-Mn}$ value is just equal to the critical distance needed for non-collinear ferromagnetism within the c-plane. It is reported that $GdMn_2Ge_2$ has a Neel temperature ($T_N$) of 365 K [11,12], below which the intra-layer Mn-Mn coupling is canted and the inter-layer coupling is antiferromagnetic. Therefore, the magnetic nature of the Mn sublattice is non-collinear antiferromagnetic below $T_N$. As a result, the net molecular field seen at the Gd site is quite small and hence the Gd moments are disordered below $T_N$. As the temperature decreases, the rare earth ordering increases and the molecular field associated with it also increases. This results in a forced ferromagnetic state for the Mn sublattice and a net non-collinear ferrimagnetic coupling between Gd and Mn moments. This transition, which is of first order in nature, occurs at $T_1=96$ K.

On the other hand, the scenario is quite different in $SmMn_2Ge_2$. As can be seen from Table l, the lattice parameters and the bond length $d^a_{Mn-Mn}$ are more than that of $GdMn_2Ge_2$. Since $d^a_{Mn-Mn}$ is more than the critical value, the inter layer Mn coupling is canted ferromagnetic in nature below the ordering temperature. It has been reported that the Curie temperature of Mn sublattice in the case of $SmMn_2Ge_2$ is about 350 K [19, 20].

As the temperature is reduced, due to the thermal contraction of the $d^{a}_{Mn-Mn}$, a first order transition from a ferromagnetic to antiferromagnetic transition within the Mn sublattice occurs at $T_2$. Reducing the temperature further causes the ordering of Sm moments which couple ferromagnetically and leads to a ferromagnetic coupling of the Mn moments. This gives rise to the transition at $T_1$, which is also first order in nature. This gives rise to the re-entrant ferromagnetism in this compound [19,20].

On substituting Sm partially for Gd, it is found that the temperature dependence of magnetization behavior is the same as in $SmMn_2Ge_2$, as is evident from Fig. 1. It may also be noted from Fig. 1 that both the FC and ZFC magnetization tend to take very low values low temperatures. The reduction in the ZFC magnetization (with decrease in temperature) reflects the domain wall pinning effect. On the other hand, the decrease in the FC magnetization suggests that there is a tendency for compensation of the magnetization of the R and Mn sublattices. Because of the fact that the Gd-Mn coupling is ferromagnetic while Sm-Mn coupling is ferromagnetic, the magnetic structure in the compounds with x=0.4 and 0.6 is quite complex.

Fig.2 shows the M-H isotherms of all the compounds at various temperatures. In the case of $GdMn_2Ge_2$, the M-H curves below $T_1$ reflect the ferromagnetic behavior, while those above $T_1$ show the antiferromagnetic behavior. These observations are consistent with the M-T plots of Fig.1. In comparison to $GdMn_2Ge_2$, the M-H isotherms of $SmMn_2Ge_2$ are more interesting. At temperatures below $T_1$, the magnetization is found to increase slightly with increase in temperature, suggesting the presence of some antiferromagnetic component in this temperature regime. In the temperature range $T_1<T<T_2$, the magnetization behavior indicates the antiferromagnetic nature. Furthermore, the saturation magnetization is almost insensitive to the temperature. In the case of the compound with x=0.4, a metamagnetic transition from the antiferromagnetic state to a forced ferromagnetic state is seen below $T_1$ and also in the regime of $T_1<T<T_2$. On the other hand, in the compound with x=0.6, a metamagnetic transition could be observed only for $T_1<T<T_2$. The absence of the metamagnetic transition below $T_1$ in this compound reflects the fact that the compound is predominantly ferromagnetic in that temperature regime. On the other hand, due to the heavy rare earth nature of Gd, the ferrimagnetic coupling will be predominant in the compound with x=0.4, which gives rise to the additional metamagnetic transition below $T_1$.

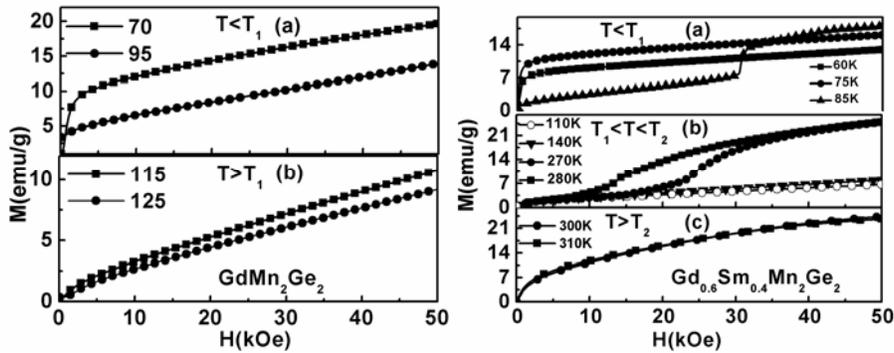

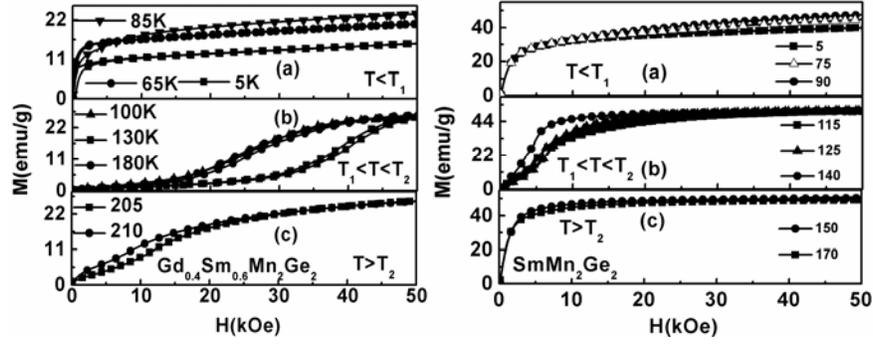

Fig.2 M-H isotherms of $Gd_{1-x}Sm_xMn_2Ge_2$ compounds at various temperatures.

The magnetocaloric effect in these compounds has been measured in terms of isothermal magnetic entropy change ($-\Delta S_M$) for various temperatures and applied magnetic fields, using the Maxwell's equation [3,5,21,22]. Fig.3 shows the temperature variation of isothermal magnetic entropy change for all the compounds, for different fields. It can be seen that in all the compounds, the entropy change is negative (positive MCE) for the transition near $T_1$ whereas, it is positive (negative MCE) for the transition at $T_2$. Moreover, the magnitude of the entropy change at $T_1$ increases with field in all the compounds, except $SmMn_2Ge_2$. The magnitude of entropy changes at $T_1$ is found to be the smallest in $SmMn_2Ge_2$ and also it decreases with increase in field. To the best of our knowledge, MCE studies on this compound have been carried out only upto a maximum field of 10 kOe and there are no reports of field dependence of MCE in the literature [23].

The peculiar MCE behavior of $SmMn_2Ge_2$ may be attributed to its special magnetic structure. Barla et. al. [19] have reported that the ground state of $Sm^{3+}$ ions in $SmMn_2Ge_2$ is $|5/2\rangle$ and that the exchange is much larger than the crystal field interaction which favors the ground state to be $|1/2\rangle$. This may be the reason for the small entropy change. The theoretical entropy change $R\ ln\ (2J+1)$ [where $R$ is the molar gas constant and $J$ is the total angular momentum quantum number] is considerably smaller for $|1/2\rangle$, as compared to that for $|5/2\rangle$. One must also be aware that $Sm^{3+}$ is a special case among the trivalent lanthanides in the sense that there is considerable mixing of the excited states with the ground state. Population of Zeeman levels by the applied field would cause an increase in the entropy. When the field is increased, the magnetic entropy increases due to this admixture effect, thereby causing the entropy change to decrease. A similar result has been observed in $PrNi_5$ [24]. The $\Delta S_M$ values obtained close to the transition at $T_1$ for a field change of 50 kOe are 6, 5.4, 3, -0.1 J/kg K in the compounds with x=0, 0.4, 0.6 and 1 respectively.

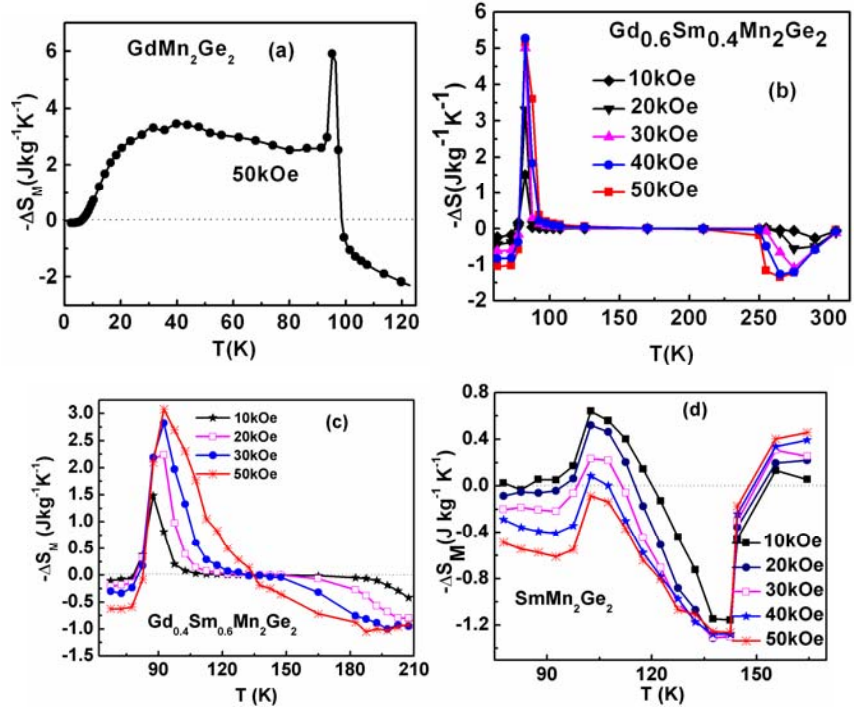

Fig.3 Temperature variation of isothermal magnetic entropy change in $Gd_{1-x}Sm_xMn_2Ge_2$ compounds in different fields.

Another interesting feature common to all the Sm-containing compounds is the positive entropy change (negative MCE) at temperatures below $T_1$. It is of interest to note that this observation is consistent with the M-H data below $T_1$ obtained in these three compounds, which shows that the magnetization increases with increase in temperature. Koyama et al. [23] have reported that the magnetic entropy change in $SmMn_2Ge_2$ for a field of 10 kOe is almost zero below 100 K. Therefore, we feel that the sign change (below $T_1$) observed in the present case occurs at fields as high as 50 kOe. This suggests that, below $T_1$, the applied field causes an increase in the magnetic entropy in the Sm-containing compounds. Chaudhary et al. [20] have reported that for fields lower than 50 kOe, the ZFC magnetization of $SmMn_2Ge_2$ is less than the FC magnetization, for temperatures lower than $T_1$. However, quite strangely, the trend reverses for a field of 50 kOe. These authors have mentioned that this observation is anomalous and the origin was not clear. We feel that the present MCE variation also corroborates with the observations of Chaudhary et al. Therefore, it is possible that the mixing of the crystal field levels mentioned earlier may be responsible for the MCE variation below $T_1$, in the case of Sm-containing compounds. Since Gd ion does not experience any crystal field effect, this behavior could be ascribed to the Sm only in all the three compounds. However, as the M-H data below $T_1$ suggests, there seems to exist an antiferromagnetic component, though much weaker than the ferromagnetic component even below $T_1$. Therefore, there may be a partial contribution from the antiferromagnetic part towards the negative MCE below $T_1$.

Fig. 3 also shows that the MCE associated with the transition at $T_2$ remains almost unchanged with Sm substitution. The maximum entropy changes are -1, -1.1 and -1.3

J/kg K for a field change of 50 kOe. Since the rare earth sublattice magnetic entropy is almost completely released well below $T_2$, the entropy change at $T_2$ could be completely associated with the Mn sublattice.

In conclusion, we find that the magnetic and magnetocaloric properties of $Gd_{1-x}Sm_xMn_2Ge_2$ correlate very well with each other. Admixture effect of crystal field levels of $Sm^{3+}$ ion may be mainly responsible for the unusual MCE behavior of the Sm-containing compounds. The temperature variations of magnetization and magnetic entropy change also suggest that the magnetic ordering below the re-entrant transition is not purely ferromagnetic, as reported. The magnetocaloric effect is found to decrease with Sm concentration due to the lower magnetic moment of Sm as compared to that of Gd.

One of the authors (KGS) thanks ISRO, Govt. of India for supporting this work through a sponsored research grant.